\documentclass[12pt,preprint]{aastex}

\begin{document}

\title{Circumbinary Planets Orbiting the Rapidly Pulsating Subdwarf B-type binary NY Vir}

\author{Qian S.-B.\altaffilmark{1,2,3}, Zhu L.-Y.\altaffilmark{1,2}, Dai Z.-B.\altaffilmark{1,2}, Fern\'{a}ndez Laj\'{u}s, E.\altaffilmark{4, 5},
Xiang F.-Y.\altaffilmark{6}, and He, J.-J.\altaffilmark{1,2}}

\altaffiltext{1}{National Astronomical Observatories/Yunnan
Observatory, Chinese Academy of Sciences (CAS), P.O. Box 110, 650011
Kunming, P.R. China (e-mail: qsb@ynao.ac.cn)}

\altaffiltext{2}{Key laboratory of the structure and evolution
celestial bodies, Chinese Academy of Sciences, P. O. Box 110, 650011
Kunming, P. R. China}

\altaffiltext{3}{Graduate University of the Chinese Academy of
Sciences, Yuquan Road 19\#, Sijingshang Block, 100049 Beijing, P. R.
China}

\altaffiltext{4}{Facultad de Ciencias Astron\'{o}micas y
Geof\'{i}sicas, Universidad Nacional de La Plata, 1900 La Plata,
Buenos Aires, Argentina}

\altaffiltext{5}{Instituto de Astrofisica de La Plata (CCT La plata
- CONICET/UNLP), Argentina}
\begin{abstract}

\altaffiltext{6}{Physics Department, Xiangtan University, 411105
Xiangtan, Funan Province, P. R. China}

We report here the tentative discovery of a Jovian planet in orbit
around the rapidly pulsating subdwarf B-type (sdB-type) eclipsing
binary NY Vir. By using new determined eclipse times together with those
collected from the literature, we detect that the
observed-calculated (O-C) curve of NY Vir shows a small-amplitude
cyclic variation with a period of 7.9\,years and a semiamplitude of
6.1\,s, while it undergoes a downward parabolic change (revealing a
period decrease at a rate of $\dot{P}=-9.2\times{10^{-12}}$). The
periodic variation was analyzed for the light-travel time effect via
the presence of a third body. The mass of the tertiary companion was
determined to be $M_3\sin{i^{\prime}}=2.3(\pm0.3)$\,$M_{Jupiter}$
when a total mass of $0.60$\,$M_{\odot}$ for NY Vir is adopted. This
suggests that it is most probably a giant circumbinary planet
orbiting NY Vir at a distance of about 3.3 astronomical units (AU).
Since the rate of period decrease can not be explained by true
angular momentum loss caused by gravitational radiation or/and
magnetic braking, the observed downward parabolic change in the O-C
diagram may be only a part of a long-period (longer than 15 years)
cyclic variation, which may reveal the presence of another Jovian
planet ($\sim2.5$$M_{Jupiter}$) in the system.

\end{abstract}


\section{Introduction}

HW Vir-like eclipsing binaries are a group of detached binary
systems that consists a very hot subdwarf B-type (sdB) primary and a
fully convective M-type secondary with periods shorter than 4 hours
(e.g., Menzies \& Marang 1986; Kilkenny et al. 1998; Drechsel et al.
2001; Ostensen et al. 2007; Polubek et al. 2007; Wils et al. 2007;
For et al. 2010; Ostensen et al. 2010). The hot sdB components in
this group of systems are on the extreme horizontal branch (EHB) of
the Hertzsprung-Russell diagram burning helium in their cores and
and having very thin hydrogen envelopes (e.g., Heber 2009).
Theoretical investigations have shown that they are formed through a
common-envelope evolution (e.g., Han et al. 2003), and will evolve
into normal cataclysmic variables (CV) (e.g., Shimansky et al.
2006). The discovery of circumbinary substellar objects orbiting
these peculiar binaries has very important implications of several
outstanding problems in astrophysics, e.g., the formation of sdB
stars and the fates of low-mass companion systems. To date, three
substellar companions were found in HW Vir and HS0705+6700 systems
as circumbinary brown dwarfs and planets (Lee et al. 2009; Qian et
al. 2009).

As for substellar objects in orbits around single sdB-type stars,
Silvotti et al. (2007) discovered a giant planet with a mass of
$3.2\,M_{Jupiter}$ and an orbital separation of 1.7\,AU, around the
hot sdB star V391 Peg. Geier et al. (2009) later announced the
detection of a close substellar companion to the hot sdB star
HD\,149382 with a short period of 2.391\,days. However, Jacobs et
al. (2011) found no evidence for the presence of the claimed
substellar object by analyzing He lines. Recently, Geier et al.
(2011) detected a brown dwarf companion to the hot subdwarf SDSS
J083053.53+000843.4 with a very short period of 0.096\,days and a
mass of 0.045-0.067\,$M_{\odot}$. Here we report the tentative
discovery of a Jovian planet with a mass of $\sim2.3\,M_{Jupiter}$
around the HW Vir-like binary star NY Vir. The planet has the lowest
mass among the substellar objects discovered before orbiting sdB
stars. Moreover, evidence may show that there is another
circumbinary planet companion in the system.

\section{New observations and changes of the O-C diagram}

\begin{figure}
\begin{center}
\includegraphics[angle=0,scale=1.1]{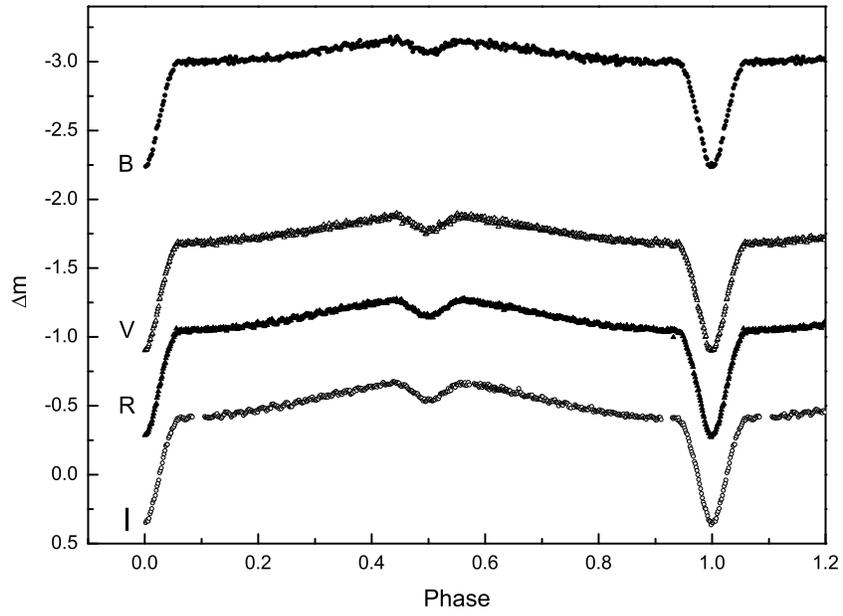}
\caption{The light curves of NY Vir in B, V, R, and I bands observed
in the time interval from May 28 to 31, 2011. Phases of the
observations were calculated with the linear ephemeris: $Min. I =
HJD\,2455711.65991+0.101015999\times{E}$.}
\end{center}
\end{figure}

NY Vir (=PG\,1336-018) was discovered as a sdB star by the
Palomar-Green survey (Green et al. 1986) and was later found to be a
HW Vir-like eclipsing binary by Kilkenny et al. (1998). It is a
short-period detached system with a period between 2 and 3 hours.
The sdB primary is also a rapid pulsator and the secondary is an
$M_5$-type star. The detailed photometric and spectroscopic
investigation was carried out by Kilkenny et al. (1998) who
determined a mass ratio of 0.3. They also found effective
temperatures $T_1=33000$\,K and $T_2=3000$\,K. A multisite (WET)
campaign identified 28 pulsation frequencies in the sdB star and
found that the amplitudes of at least the strongest frequencies were
varying on timescales of days (Kilkenny et al. 2003).

\begin{table*}
\caption{New CCD times of light minimum of NY Vir.}
\begin{center}
\begin{tabular}{llllll}\hline
BJD (days)  & Errors (days) & Min. & Filters & E &
Telescopes\\\hline\hline
2454211.16991 & 0.00007 & I & R & 39477   & The 1.0-m telescope\\
2455709.64037 & 0.00001 & I & I & 54311   & The 2.15-m telescope\\
2455709.69102 & 0.00004 & II& I & 54311.5 & The 2.15-m telescope\\
2455711.55967 & 0.00001 & I & R & 54330   & The 2.15-m telescope\\
2455711.61016 & 0.00004 & II& R & 54330.5 & The 2.15-m telescope\\
2455711.66070 & 0.00001 & I & V & 54331   & The 2.15-m telescope\\
2455711.71125 & 0.00006 & II& V & 54331.5 & The 2.15-m telescope\\
2455712.56984 & 0.00002 & I & B & 54340   & The 2.15-m telescope\\
2455712.62042 & 0.00005 & II& B & 54340.5 & The 2.15-m telescope\\
\hline\hline
\end{tabular}
\end{center}
\end{table*}

\begin{figure}
\begin{center}
\includegraphics[angle=0,scale=1.2]{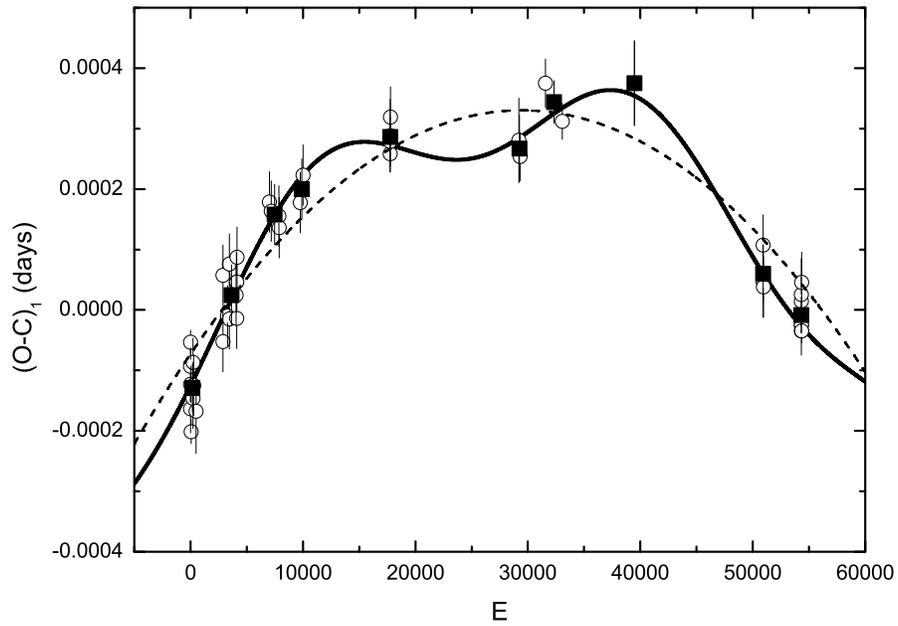}
\caption{The O-C diagram of the sdB-type eclipsing binary NY Vir
respect to Eq. (3). The solid line refers to the best fit of the O-C
curve by a combination of a periodic variation with a small
amplitude of 6.1\,s and a period of 7.9\,years and a downward
parabolic change (the dashed line). Open circles refer to the
original data points, while solid squares to their averaged values.
}
\end{center}
\end{figure}

38 times of light minimum of NY Vir were published by using the
telescopes in South African Astronomical Observatory (SAAO)
(Kilkenny et al. 1998, 2000, 2011). Linear ephemerides,
\begin{equation}
Min. I (HJD) = 2450223.36142 + 0.1010174 \times E
\end{equation}
and
\begin{equation}
Min. I (HJD) = 2450223.36134 + 0.101015999 \times E
\end{equation}
were determined by Kilkenny et al. (1998 and 2000), where E is the
cycle number. A recent period investigation of NY Vir (Kilkenny et
al. 2011) suggests that its period is decreasing at a rate of
$\dot{P}=-11.2\times{10^{-13}}$\,d per orbit. The mechanism that
causes this decrease is unknown. To get more data for this binary
star, it was observed with the 1.0-m telescope in Yunnan observatory
on April 20, 2007. Then, after read the paper by Kilkenny et al.
(2011), we monitored it again in the time interval from May
28 to 31, 2011 with the 2.15-m "Jorge Sahade" telescope in
Argentina. During the observation a Versarray 1300B CCD camera was
used. The complete light curves in B, V, R, and I bands were
obtained and are displayed in Fig. 1. By using those photometric
data, 9 eclipse times were determined. In order to obtain a high
data precision, we applied a standard time system Barycentric Julian
Dynamical Date (BJD), which are converted by the code of Stumpff
(1980), to all eclipse timings and orbital period analysis. The
derived eclipse times in BJD are listed in Table 1.

With all of available times of light minimum, the linear ephemeris
published by Kilkenny et al. (1998 and 2000) was revised as,
\begin{equation}
Min. I (BJD) = 2450223.362213(8) + 0.1010159673(2) \times E,
\end{equation}
where BJD\,2450223.362213 is the initial epoch, and 0.1010159673 is
the revised period. The variance is derived as
$1^{d}.7\times10^{-4}$. The $(O-C)_{1}$ values respect to the linear
ephemeris were calculated. The corresponding $(O-C)_{1}$ diagram
is plotted in Fig. 2. As shown in the figure, a downward parabolic fit
(dashed line in Fig. 2) can not describe the general trend of the
$(O-C)_{1}$ curve good. This indicates that, to describe the general
$(O-C)_1$ trend satisfactorily, a combination of a downward
parabolic change and a cyclic oscillation is required. Using the
least squares method, we determined,
\begin{eqnarray}
(O-C)_{1}&=&-0.000073(\pm0.00002)+2.7(\pm0.2)\times{10^{-8}}\times{E}\nonumber\\
     &&-4.6(\pm0.4)\times{10^{-13}}\times{E^{2}}\nonumber\\
    & &+0.000071(\pm0.000017)\sin[0.^{\circ}0126(\pm0.0007)\times{E}+312.^{\circ}1(\pm20.^{\circ}1)],
\end{eqnarray}
with a variance of $3^{d}.8\times10^{-5}$ and the
$\chi^{2}\approx1.2$. Using the F-test as discussed by Pringle
(1975) to assess the significance of quadratic and sinusoidal terms
in Eq. 4, the parameter $\lambda$ can be derived to be $180$, which
indicates that it is significant well above $99.99\%$ level. The
rate of the orbital period decrease is close to that determined by
Kilkenny et al. (2011). Considering that the influence of the
pulsations on the minima time measurements is a plausible reason for
the observed season scatters shown in Fig. 2 and 3, we average over
the eclipse times within one year as a window of time. The averaged
points are well obey our fitting curve in a long base line. Thus, we
believed that the variation in O-C diagram should be real.

After the downward parabolic variation was
removed, the $(O-C)_2$ curve is plotted in upper panel of Fig. 3 where a cyclic change
is seen more clearly. The cyclic variation has an amplitude of 6.1
seconds and a period of 7.9 years. As pointed out by Kilkenny et al.
(2011), eclipsing times for NY Vir are
less accurate than for AA Dor because of the pulsating. However, errors introduced should be of
the order of 0.00005\,d or less (Kilkenny et al. 2001) and
essentially random. The mean values of $(O-C)_2$ suggest that the
systematically cyclic variation in the $(O-C)_2$ diagram is true.
After all of the changes were removed, the residuals are shown in
the lower panel of Fig. 3. It can be seen in the panel that the
scatter of the residuals is in a random way indicating the scatter
may be caused by the pulsating of the sdB star. Moreover, no
systematic changes can be traced there, which suggests that Eq. (4)
can give a good fit to the observations.

\begin{figure}
\begin{center}
\includegraphics[angle=0,scale=1.1]{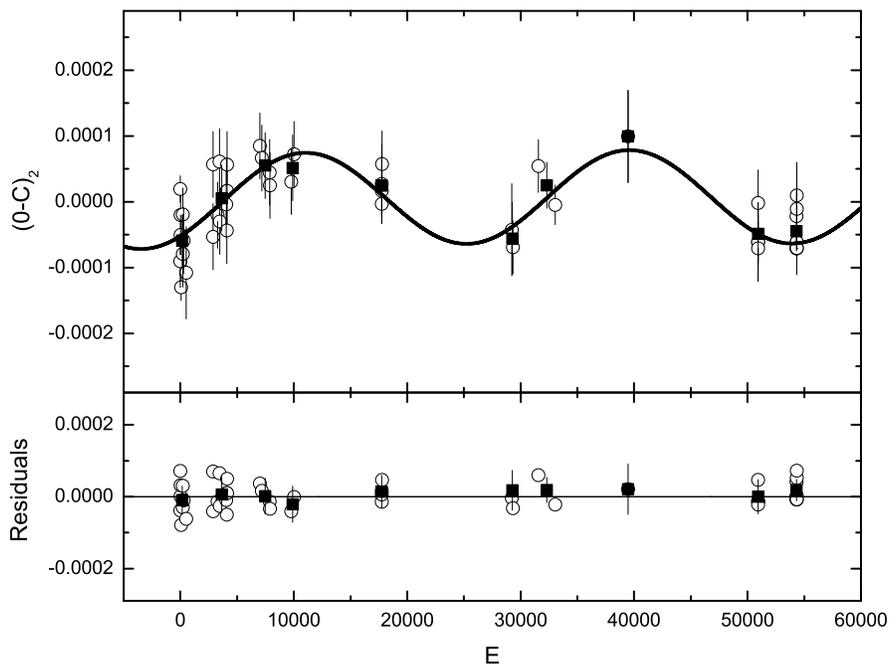}
\caption{The cyclic variation of NY Vir (the light travel-time
effect via the presence of the giant circumbinary planet) after the
downward parabolic change is subtracted from the $(O-C)_1$ diagram.
Symbols are the same as those in Fig. 2. The mean values follow the
general $(O-C)_2$ trend, which suggests that the cyclic variation is
real.}
\end{center}
\end{figure}

\section{Discussions and conclusions}

Investigations of several authors (e.g., Qian et al. 2008;
Brinkworth et al. 2005) showed that the magnetic activity cycles of
a fully convective component star (i. e., the Applegate mechanism)
(Applegate 1992) can not explain the observed cyclic variations in
the O-C diagrams of close binaries because of the problem of energy.
Moreover, a recently statistical study by Liao and Qian (2010)
suggests that light travel-time effect is the most probable
mechanism to cause cyclic changes in the O-C diagrams of close
binary stars. Therefore, we analyzed NY Vir for the light
travel-time effect that arises from the gravitational influence of a
tertiary companion. The possible presence of a companion object produces the
relative distance changes of the eclipsing pair as it orbits the
barycenter of the triple system.

By considering a typical mass of $M_{1}=0.46M_{\odot}$ for sdB-type
stars and using the mass ratio of $q=0.3$ determined by Kilkenny et
al. (1998), the mass of the secondary is estimated as
$M_{2}=0.14M_{\odot}$. Then, the mass function and the mass of the possible
tertiary companion were derived by using the following equation,
\begin{equation}
f(m)=4\pi^{2}(a_{12}\sin{i^{\prime}})^{3}/GT^{2}=(M_{3}\sin{i^{\prime}})^{3}/(M_{1}+M_{2}+M_{3})^{2},
\end{equation}
where $M_{3}$ is the mass of the third body, G is the gravitational
constant, and T is the period of the $(O-C)_2$ oscillation.
$a_{12}\sin{i^{\prime}}$ can be determined by,
\begin{equation}
a_{12}\sin{i^{\prime}}=A\times{c}
\end{equation}
where A is the semi-amplitude of the $(O-C)_2$ oscillation and $c$
the speed of light. The results are:
$f(m)=3.0(\pm0.4)\times{10^{-8}}\,M_{\odot}$ and
$M_3\sin{i^{\prime}}=2.3(\pm0.3)$\,$M_{Jupiter}$. When the orbital
inclination of the third body is larger than $9.6^{\circ}$, the mass
of the tertiary component corresponds to
$2.3\,M_{Jupiter}\le{M_3}\le14\,M_{Jupiter}$. In this situation, as
a few good cases have been made for circumbinary planets (e.g., Lee
et al. 2009; Beuermann et al., 2010; Qian et al. 2010, 2011), it
should be a planetary object. The orbital separation $d_3$ of the
Jupiter-like planet candidate is about $3.3(\pm0.8)$\,AU.

The downward parabolic change in the O-C diagram (the dashed line in
Fig. 2) implies a period decrease at a rate of
$\dot{P}=-9.2\times{10^{-12}}$ (or 1s in about 3,447 years). This
corresponds to $\dot{P}=-3.36\times{10^{-9}}$\,days/year, which is
close to the value obtained in literature (Kilkenny et al. 2011).
The orbital separation of the two component stars in NY Vir is
estimated as $0.77\,R_{\odot}$. By using the equation given by Kraft
et al. (1961) and Faulkner (1971),
\begin{equation}
\frac{\dot{P}_{GR}}{P}=-3\frac{32G^3}{5c^5}\frac{M_1M_2(M_1+M_2)}{d^4},
\end{equation}
where $P$ is the orbital period and $d$ the distance between both
components, the contribution of gravitational radiation (GR) to the
period decrease can be computed as
$\dot{P}_{GR}=-0.026\times{10^{-9}}$\,days/year. This is about two
orders smaller than the observed value indicating that it can not be
caused by GR. This decrease can be explained by magnetic braking
(MB) as in the case of HW Vir (Qian et al. 2008; Lee et al. 2009).
However, it is more widely accepted that MB is stopped for fully
convective stars (Rappaport et al. 1983; Spruit \& Ritter 1983).

Therefore, as in the case of HU Aqr, the more plausible reason for
the observed long-term period decrease is that it is only a part of
a long-period cyclic variation (longer than 15 years), revealing the
possible presence of another planet in the planetary system. This is
the same as that of HU Aqr where a planetary system is orbiting the
eclipsing polar (Qian et al. 2011). We estimate that the mass of the
second planet is about $2.5$\,$M_{Jupiter}$. Some investigators
(e.g., Horner et al. 2011) think that the serious problem for the
presence of the fourth body is the dynamical stability of the
circumbinary planetary system. However, the 3D dynamic-aware
analysis of the stability of circumbinary orbits by Doolin et al.
(2011) reveals that these orbits are surprisingly stable throughout
binary mass fraction -- eccentricity parameter space.

If the cyclic variations in the O-C diagram are caused by the
light-travel time effects via the presence of planetary objects,
they should be strictly periodic. To check the existence of the
planetary system, more eclipse times are needed in the future.
Moreover, At the maximum point of the O-C curve, the central
eclipsing binary NY Vir is at the farthest position of the orbit,
while the circumbinary planet is closest to the observer. If the
tertiary companion is coplanar to the central binary, the
circumbinary planet should be in the light of the eclipsing binary
and transit the binary component stars. Therefore, searching for the
transits of the binary components by the circumbinary planets at the
O-C maxima can ascertain the planetary system. More recently, a
Saturn-like planet transits an M-type eclipsing binary was reported
by Doyle et al. (2011), which provides the first direct evidence for
the presence of a circumbinary planet.

As in the cases of other HW Vir-like binary stars, NY Vir was
evolved through a common envelope (CE) after the more massive
component star in the original system evolves into a red giant. The
ejection of CE removed a large amount of the angular momentum, and
the present EHB star with a low-mass stellar companion in a
short-period binary was formed. The distance between the central EHB
binary and the circumbinary planet is about 3.3\,AU. By assuming the
mass of the main-sequence (MS) progenitor of the EHB star is
$M_{MS}=1.0$\,$M_{\odot}$, the orbital separation of the planet
around the MS progenitor is estimated as (e.g., Bear \& Soker 2011):
$a_0\cong{(M_{EH}+M_2)/(M_{MS}+M_2)}\approx1.74$\,AU (by ignoring
tidal interaction). When the primary star evolved to a red giant
branch star before the system entered a CE, it has a reached radius
of $\sim{100}$\,$R_{\odot}\sim{0.5}$\,AU. Therefore, the secondary
should be at a distance of a little large, say 0.8 AU. Can a system
of two stars ($M_1=1.0\,M_{\odot}$ and $M_2=0.14\,M_{\odot}$) at an
orbital separation of 0.8\,AU have a planet at 1.74\,AU in a
stable orbit? The 3D numerical simulations by Doolin et al.
(2011) indicates that the orbit is stable. A system with two M-type stars ($M_1=0.69\,M_{\odot}$
and $M_2=0.20\,M_{\odot}$) at an orbital separation of 0.22\,AU was
reported to have a Saturn-like planetary companion at a distance of
0.7\,AU (Doyle et al. 2011). If the detection of a planetary
companion to NY Vir at orbital separation of 3.3\,AU is true, it can
give some constraints on the stellar evolution and the interaction
between red giants and their planetary companions. Moreover, it is
interesting to pointed out that the orbital separations between the
central binaries and the inner substellar objects in the three
systems, i.e., HW Vir (Lee et al. 2009), HS\,0705+6700 (Qian et al.
2009), and NY Vir, are all in a small range of 3.3-3.6\,AU. It is
unclear whether these agreements are fortuitous or there are
physical reasons.

\acknowledgments{This work is partly supported by Chinese Natural
Science Foundation (No. 11133007, No.11003040, and No.10878012) and
by the West Light Foundation of the Chinese Academy of Sciences. New
CCD photometric observations of NY Vir were obtained with the 1.0-cm
telescope in Yunnan observatory and with the 2.15-m "Jorge Sahade"
telescope.}

\end{document}